\documentclass[aps, prl,preprint,showpacs,superscriptaddress]{revtex4}
\usepackage{amssymb}
\usepackage{amsthm,epsf}
\usepackage{amsfonts}
\usepackage{epsfig}
\usepackage{graphicx}
\usepackage{latexsym}
\begin{document}
\title {\bf Evolving Networks and Birth-and-Death Processes \footnote {This
research is supported by Hong Kong Research Grant Council for
research grants HKUST698/01E and HKUST6133/02E, and  partially by
the 863 Project through grant 2002AA234021.}}

\author{Dinghua Shi}
\affiliation{ \small Department of Mathematics, Shanghai
University, Shanghai 200436, China}
\author{Liming Liu}
\email{liulim@ust.hk}
 \affiliation{
\small Department of Industrial Engineering and Engineering Management\\
\small Hong Kong University of Science and Technology, Clear Water
Bay, Hong Kong }
\author{Xiang Zhu}
 \affiliation{
\small Department of Industrial Engineering and Engineering Management\\
\small Hong Kong University of Science and Technology, Clear Water
Bay, Hong Kong }
\author{Huijie Zhou}
\affiliation{ \small Department of Mathematics, Shanghai
University, Shanghai 200436, China}
\author{Binbin Wang}
\affiliation{ \small Department of Mathematics, Shanghai
University, Shanghai 200436, China}

%


\date{March 23, 2005}

\begin{abstract}
Using a simple model with link removals as well as link additions,
we show that an evolving network is scale free with a degree exponent in the range
of $(2,4]$. We then establish a relation between the network
evolution and a set of non-homogeneous birth-and-death processes,
and, with which, we capture the process by which the
network connectivity evolves.
We develop an effective algorithm to compute
the network degree distribution accurately.
Comparing analytical and numerical results with simulation, we
identify some interesting network properties and verify the
effectiveness of our method.
\end{abstract}

\pacs{89.75.Hc; 64.60.Fr; 87.23.Ge}

\maketitle

\bigskip
\textbf{\large Growing Networks and Pure Birth Processes}

The first growing network model, i.e., the BA model proposed by
Albert, Baraba$\acute{a}$si, Jeong \cite{Barabasi99a}, predicts a
power-law network degree distribution with exponent $\gamma=3$,
whereas the degree exponents of many real complex networks are
found empirically in the range of $(2,4)$
\cite{Newman03,{Albert}}. This has motivated extensive research to
modify the basic BA model to match with practical scale-free
networks. These variations can be summarized by the following general BA model:

(i) Initialization: $n_0$ nodes are given at time $t=0$;

(ii) Growth: at the $t$th time step, a new node and $m(t)(\leq
t+n_0)$ new links from this node are added;

(iii) Preferential attachment: the new node is connected to an
existing node $i$ according to the following probability
$\Pi(k_i)=f(k_i)/\sum_j f(k_j)$, where $k_i$ is the number of
degrees of node $i$ and $f(k_i)$ is pre-selected function.

With general $m(t)$ and $f(k_i)$, this
model is analytically intractable and is very complicated for simulation study. 
We may simplify this model in two ways. Setting $m(t)=m$,
i.e., the number of links in the network increase linearly,
the growth of the network is stationary.
Assuming $f(k_i)=k_i$ further, we have the basic BA model. If we set
$f(k_i)=(1-p)k_i+p$ instead, where $p$ is the probability of
randomly selecting an old node, the model reduces to Liu et al.
\cite{Liu02}'s model with $\gamma=3+p/[m(1-p)]$. In Bianconi and
Barabsi \cite{Bianconi01}'s fitness model, $f(k_i)=\eta_i k_i$,
where $\eta_i$ is chosen from a distribution $\rho(\eta)$. If
$\rho(\eta)$ is uniform, $\gamma=2.255$. Alternatively, we can
also first set $f(k_i)=k_i$. With a time-dependent number of new
links added at each time step, the growth of a network is
non-stationary. For example, with the accelerating function
$m(t)=mt^{\theta}$, $0\leq \theta<1$ proposed by Dorogovtseva and
Mendes \cite{Dorogovtseva01}, the degree exponent
$\gamma=(3-\theta)/(1-\theta)$ and the non-stationary exponent
$z=2\theta/(1-\theta)$. Shi, Chen and Liu \cite{Shi04} proposed a
slower accelerating function $m(t)=m\ln t$, $t\geq 2$.

\begin{figure}[htp]
 \begin{center}
  \includegraphics[width=6cm]{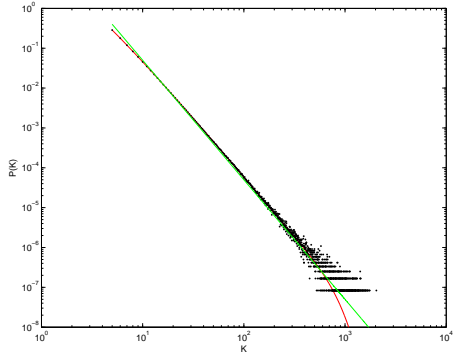}\\
 \caption{{\footnotesize The degree distribution for the BA model with $m=5$, $S=3$
and $t=150000$: by the analytical method in
solid green line; by simulation in dotted black line; and by the
PBP method in solid red line.}}\label{fig 1}
\end{center}
\end{figure}

Shi, Chen and Liu \cite{Shi04} established a relation
between the connectively of a growing network and a set of
non-homogeneous pure birth processes (PBP) and found numerically
that $\gamma\approx 3.1$ and the non-stationary exponent is very
small for the case of $m(t)=m\ln t$. It can be observed that the degree
distribution curves of growing networks in \cite{Shi04} is {\it
snake} like with a slightly downward bending head section as
illustrated in figure \ref{fig 1} (see, also \cite{Shi04} for similar figures).

There are other ways to extend the BA model. Readers can refer to
\cite{Albert} for a comprehensive review.

\bigskip
\textbf{\large A Model of Evolving Networks and Dynamic Equation}

Albert and Baraba$\acute{a}$si \cite{Ab} considered a model of the
evolving network in which some old links are rewired at each time
step (see Section 4). We observe from many real networks that
beside adding new nodes and links, some old nodes and links can
also be removed as a network evolves. In other words, many
networks display a dynamic evolving process. 
We propose the following simple model to capture the basic features of the above evolving network.

(i) Initialization: There are $n_0$ fully connected initial nodes.

(ii) Link removal: At each time step, $c$ old links are removed
as follows. We first select node $i$ with the anti-preferential
probability similar to that used in \cite{K02}

\begin{equation}
\label{eq: antip} \Pi^*(k_i)=ak_i^{-1},
\end{equation}
where $a$ is used as a normalized factor such that
$a^{-1}=\sum_ik_i^{-1}$.
We then choose node $j$ from the neighborhood of node $i$ (denoted
by $O_i$ ) with probability $K_i^{-1}\Pi^*(k_j)$, where
$K_i=\sum_{j\in O_i }\Pi^*(k_j)$. The link connecting nodes $i$
and $j$ is removed. We repeat this procedure $c$ times to remove
$c$ existing links. Finally, isolated nodes are removed from the network.

(iii) Link addition: At each time step, a new node is added to the
system and $m(\leq n_0)$ new links from the new node are connected
to $m$ different existing nodes. A node $i$ with degree $k_i$ will
receive a connection from the new node with a Bayes' preferential probability
\begin{equation}
\label{eq:prep} \Pi(k_{i}) = \frac{k_{i}+1}{\sum_{j}(k_{j}+1)}.
\end{equation}

The above model is different from the existing ones in that 
it removes links instead of rewiring links. 
Furthermore, isolated nodes are removed in our model.
 
By the continuum theory, $k_i(t)$ approximately satisfies the
following dynamic equation
\begin{eqnarray}
\frac{\partial k_i}{\partial t}&=&m\Pi(k_i)-c\left[\Pi^*(k_i)+
\sum_{j\in O_i}\Pi^*(k_j)K_j^{-1}\Pi^*(k_i)\right] \nonumber\\
\label{eq: dk} &\approx&m\frac{k_{i}+1}{[2(m-c)+1]t}-c\frac{2}{t},
\end{eqnarray}
where the last approximation is based on $\sum_j
(k_j+1)=2(m-c)t+N(t-1) \approx [2(m-c)+1]t$, $\sum_{j\in
O_i}K_j^{-1}\Pi^*(k_j) \approx 1$, and, in the mean-field sense,
$ak_i^{-1} \approx [N(t-1)]^{-1} \approx 1/t$, in which $N(t-1)$
is the number of non-isolated nodes at time step $t$.

Let $t_i$ be the time step when node $i$ is added to the network.
Initially, node $i$ has $k_i(t_i) = m$ links, thus the above
equation has the following solution 
\begin{equation}
 \label{eq: so}
 x_{i}(t)=|k_{i}(t)+B-m|= B\left ( \frac{t}{t_i} \right )^\beta ,
\end{equation}
with the dynamic exponent
\begin{equation}
\label{eq: de} \beta=\beta(m,c)=\frac{m}{2(m-c)+1}
\end{equation}
and the dynamic coefficient
\begin{equation}
\label{eq: co} B=B(m,c)=m+\frac{m-2c[2(m-c)+1]}{m}.
\end{equation}

In the solution procedure, we require $0<\beta<1$ and $B>0$ for
the solution to be feasible. Some simple analysis of the above
formulas shows that $m>2c$ is a sufficient condition for equation
(\ref{eq: dk}) to have a feasible solution.

Assume that $t_i$ follows a uniform distribution over interval
$(0, t)$. We have, by (\ref{eq: so})
\begin{eqnarray}
P(x)=\frac{1}{\beta}B^{1/\beta} x^{-\gamma},
\end{eqnarray}
where $x\in [B, \infty)$ following from (\ref{eq: so}) and the
degree exponent
\begin{equation}\label{eq: se}
\gamma=1+\frac{1}{\beta}=3+\frac{1-2c}{m}.
\end{equation}
Equation (\ref{eq: se}) shows that this system
self-organizes into a scale-free network with $2<\gamma \leq 4$.

The next step is to obtain the network degree distribution. For $B
\geq m$, we have, following the standard mean field approach
\cite{Ab}, the explicit solution of dynamic equation (\ref{eq:
dk})
\begin{equation}
\label{eq: so1} k_i(t)=B \left [ \left ( \frac{t}{t_i}
\right)^\beta -1 \right ] + m
\end{equation}
and the network degree distribution
\begin{equation}
\label{eq: ap} P(k) = \frac{1}{\beta}B^{1/\beta}(k+B-m)^{-\gamma}.
\end{equation}

For $B<m$, the continuum theory does not render an accurate
solution, and we need a different method. 

\bigskip
\textbf{\large Birth-and-Death Processes of Network Connectively}

The dynamics of the degree of a node in an evolving network is
closely related to Markov processes. Let $K_i(t)$ be the degrees
of node $i$ at time $t$. Since $K_i(t+1)$ only depends on $K_i(t)$
and allows the removal of old links for our model, $\{K_i(t),
t=i,i+1,\ldots\}$ is a discrete-time Markov process with the state
space $\Omega=\{0,1,2,\ldots\}$.

By (\ref{eq: dk}), the probability that node $i$ with degree
$k(\geq 1)$ is connected to a new node at time step $t$ is
approximately $g_{t}(k)=(1-2c/t)m\Pi(k) \approx m\Pi(k)
=m(k+1)/([2(m-c)+1]t)$.
The probability that note $i$'s degree decreases by $1$ is
approximately $(2c/t) \left[1-m\Pi(k_i)\right] \approx 2c/t$,
while the probability that its degree decreases by more than 1 is
$o(t)$ and will be ignored. Thus, the probability that the degree
of node $i$ remains the same is $h_{t}(k)=1-g_{t}(k)-2c/t$. This
shows that $K_i(t)$ is in fact a non-homogeneous birth-and-death
process (BDP). In addition, we set $p_{00}=1$ since we remove isolated nodes
and $p_{kk} =1$ when $k>m+t-i$.
In summary, for $t=i,i+1,i+2,\ldots$, the
one-step transition probability matrix of node $i$ at time $t$ is given by
\begin{equation}
\textbf{P}_{i}(t+1)=\left[\matrix{1 & 0 && \cr
 & \cr
2c/t & h_{t}(1) & g_{t}(1) && \cr
 \ddots&\ddots & \ddots && \cr
 & \cr
 & 2c/t & h_{t}(m+t-i) & g_{t}(m+t-i) && \cr
 & \cr
 &&0&1 & 0  & \cr
 &&\hspace{1cm}\ddots&\hspace{1cm}\ddots & \hspace{1cm}\ddots &
 }  \right].
\end{equation}

Denote $f_{i,n}(t)=P\{K_i(t)=n\}$ for $n=0,1,2,\ldots$ and
$\vec{f}_i(t)=(f_{i,0}(t),f_{i,1}(t),\cdots,f_{i,n}(t),\cdots)$.
Obviously, $\vec{f}_i(i)=(0,0,\cdots,0,1,0,\cdots)=\vec{e}_m$ where $e_{i,m}(i)=1$. 
By density evolution, the $(t+1)$th-step probability vector $\vec{f}_i(t+1)$
for node $i$ is given by
\begin{equation}
\vec{f}_i(t+1)=\vec{e}_m\cdot \textbf{P}_i(i+1)\cdot
\textbf{P}_i(i+2)\cdots \textbf{P}_i(t+1), ~~~~ t=i,i+1,\ldots.
\end{equation}
Let
\begin{equation}
\label{mc dd} \vec{F}^{(S,t)}(t+1)= \sum_{i=S}^{t} \vec{f}_i (t+1)
= \sum_{i=S}^{t} \vec{e}_m \textbf{P}_i (i+1) \cdot \textbf{P}_i
(i+2) \cdot \cdot \cdot \textbf{P}_i (t+1),
\end{equation}
where the integer $S\geq 1$ is needed
technically for the transition probability matrix. For the
choice of $S$ and its impact on computation, please refer to
\cite{Shi04}. 

Generally, it would be extremely difficult to
calculate (\ref{mc dd}). 
Fortunately, we can find the following relations:
\begin{equation} \vec{e}_m \textbf{P}_i(t)=\vec{e}_m \textbf{P}_{S}(t), ~~~~ i=S+1, S+2, ...; ~~t=i+1, i+2,
...;
\end{equation}
and, in general, for 
$s=1,2, ... ~$
\begin{equation}
\vec{e}_m \textbf{P}_i (t)\textbf{P}_i (t+1) \cdot \cdot \cdot
\textbf{P}_i (t+s) = \vec{e}_m \textbf{P}_S (t)\textbf{P}_S (t+1)
\cdot \cdot \cdot \textbf{P}_1 (t+s).
\end{equation}
Thus we obtain the following key algorithm
\begin{equation}
\label{simple} \vec{F}^{(S,t)}(t+1)=((\cdot \cdot \cdot
(\vec{e}_{m}\textbf{P}_{S}(S+1)+\vec{e}_{m})\textbf{P}_{S}(S+2)+\cdot
\cdot \cdot )+\vec{e}_{m})\cdots \textbf{P}_{S}(t+1).
\end{equation}
The right-hand side of (\ref{simple}) can be efficiently computed
with a complexity of $O(t^2)$ \cite{Shi04}. 

The degree distribution of a network can be determined
by the average of the degree distributions of all the nodes. Therefore, for a
sufficiently large $t$, we have
\begin{equation}
\label{approx} P(k)\approx
P(k,t+1)=\frac{F_{k}^{(S,t)}(t+1)}{t-S+1}.
\end{equation}

As a bonus, we can also estimate the
number of non-isolated nodes from (\ref{approx}) as follows
\begin{equation}
\label{est} N(t)\approx (t+1)[1-P(0,t+1)],
\end{equation}
noting that $P(0,t+1)$ is the probability that a node is isolated
at time step $t+1$. This
index 
cannot be obtained from (\ref{eq: ap}).

\bigskip

We note that there are also inaccuracies in the
transition probability matrices, and we can only perform a finite
number of computation steps to estimate the asymptotic network
behavior. To verify the BDP 
method, we compare the computation results with simulation.

\begin{figure}[htp]
 \begin{center}
  \includegraphics[width=8.5cm]{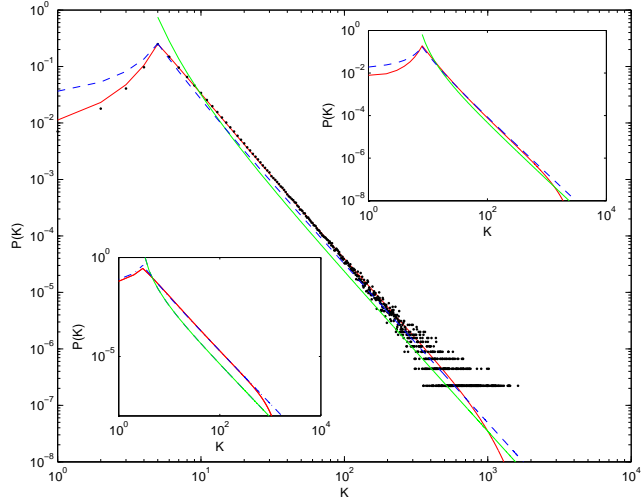}\\
 \caption{{\footnotesize The degree distribution obtained by 
(\ref{eq: ap}) in solid green line; by 
(\ref{eq: a}) in dash blue line; by simulation in dotted black
line; and by the BDP 
method in solid red line for
$m=5$, $c=1$, $S=4$ and $t=150000$.
The settings for the inserts are the same as those of the main
figure, except $m=3$ and $c=1$ for the bottom left insert and
$m=8$ and $c=2$ for the top right insert, and no simulation
results are given.}}\label{fig 2}
\end{center}
\end{figure}

From figure \ref{fig 2}, we see that the network degree
distribution curves obtained by the BDP 
method and by simulation match very well.
Our method predicts a {\it horse head}-like distribution curve,
with its middle section displaying the expected scale-free state. Because we can only
perform a finite number of computation steps,
there is an inward bend at the tail of the distribution curve
(see \cite{Shi04} for more detailed discussion). The degree
exponent and coefficient can be estimated by applying the least
square method to the data generated from $P(k)\in (10^{-4},10^{-6})$.

The numerical results from the birth-death processes clearly show that
the network degree distribution curve has a very different head section.
This motivates us to {\it construct} the following approximation for the degree distribution when $B<m$,
noticing also from (\ref{eq: so}) that, when $B<m$, $x_i(t)$ achieves the minimum at $k_i(t)=m$ and is
symmetric
\begin{equation}
\label{eq: a} P(k) \approx
\left\{%
\begin{array}{ll}
    C\frac{1}{\beta}(k-\mu)^{-\gamma},~& k\geq m\\
    C\frac{1}{\beta}(2m-\mu-k)^{-\gamma},~& k<m
\end{array}%
\right.
\end{equation}
where $\mu$ is a fitted parameter and the coefficient
\begin{equation}\label{eq: c}
C=(2m-\mu)^{\frac{1}{\beta}}/\left [ 2 \left ({\frac{2m-\mu}{m-\mu}}\right )^{\frac{1}{\beta}}-1 \right]
\end{equation}
is a normalizing constant such that $\int_0^\infty P(k)dk=1$.

Empirically, we find that when $\mu=0.2m+(c-1)$,
the approximation
(\ref{eq: a}) is very accurate for the overall distribution and
captures the pattern of the small degree distribution, as
shown in the small inserts in figure 2. 
The figures also show that (\ref{eq: ap}) cannot provide probabilities for
degrees smaller than $m$, and it visibly over estimates other
small degree probabilities and under estimate large degree probabilities.

\bigskip
\textbf{\large Application to the Albert-Barab$\acute{a}$si model} 

The model proposed by Albert Barab$\acute{a}$si \cite{Ab} starts with $n_0$ isolated
nodes, and
performs one of the following operations at each time step:

(i) Add $m(\leq n_0)$ new links with probability $p$:
Select a node randomly as the starting point of the new link and then
select the other end of the link with the preferential
probability (\ref{eq:prep}). Repeat this process $m$ times.

(ii) Rewire $m$ links with probability $q$:
Select randomly a node $i$ and a link $l_{ij}$ connected to it.
Remove this link and replace it with a new link $l_{ij'}$ that
connects $i$ to node $j'$ which is chosen with the preferential
probability (\ref{eq:prep}). Repeat this process $m$ times.

(iii) Add one new node with probability $r=1-p-q$: The new node
has $m$ new links that are connected to different existing nodes with the preferential probability
(\ref{eq:prep}).

By the continuum theory, Albert and Barab$\acute{a}$si obtained
the following dynamic equation
\begin{eqnarray}
\label{eq: abd} \frac{\partial k_i}{\partial t}\approx
m\frac{k_{i}+1}{[2m(1-q)+r]t}+m\frac{p-q}{rt},
\end{eqnarray}
and from which they derived the network degree distribution
\begin{equation}
\label{eq: abp}
P(k)=\frac{1}{\beta}(m+\tau)^{\frac{1}{\beta}}(k+\tau)^{-\gamma},
\end{equation}
where $\tau =(p-q)(2m(1-q)/r+1)+1$ and the degree exponent
\begin{equation}\label{eq: ae}
\gamma=1+\frac{1}{\beta}=3+\frac{r-2mq}{m}.
\end{equation}
Obviously, (\ref{eq: abp}) is valid only when $m+\tau >0$.
Thus for the Albert-Baraba$\acute{a}$si model, the network degree distribution is scale-free
only when parameters $p$ and $q$ satisfy
\begin{equation}
\label{eq: abq}
q<q_{max}=min\{1-p, (m+1-p)/(2m+1)\}.
\end{equation}

Now, we use birth-and-death processes to discuss the Albert-Baraba$\acute{a}$si model.
By (\ref{eq: abd}), the one-step transition probability matrix of node $i$ at time $t$ is given by
\begin{equation}
\textbf{P}_{i}(t+1)=\left[\matrix{1-g_{t}(0) & g_{t}(0) && \cr
 & \cr
 mq/rt& h_{t}(1) & g_{t}(1) && \cr
 \ddots&\ddots & \ddots && \cr
 & \cr
 & mq/rt & h_{t}(m+t-i) & g_{t}(m+t-i) && \cr
 & \cr
 &&0&1 & 0  & \cr
 &&\hspace{1cm}\ddots&\hspace{1cm}\ddots & \hspace{1cm}\ddots &
 }  \right] ,
\end{equation}
where $g_{t}(k)\approx (m(k+1)/([2m(1-q)+r]t)) +mp/rt$ and
$h_{t}(k)=1-g_{t}(k)-mq/rt$.

\begin{figure}[htp]
 \begin{center}
  \includegraphics[width=6cm]{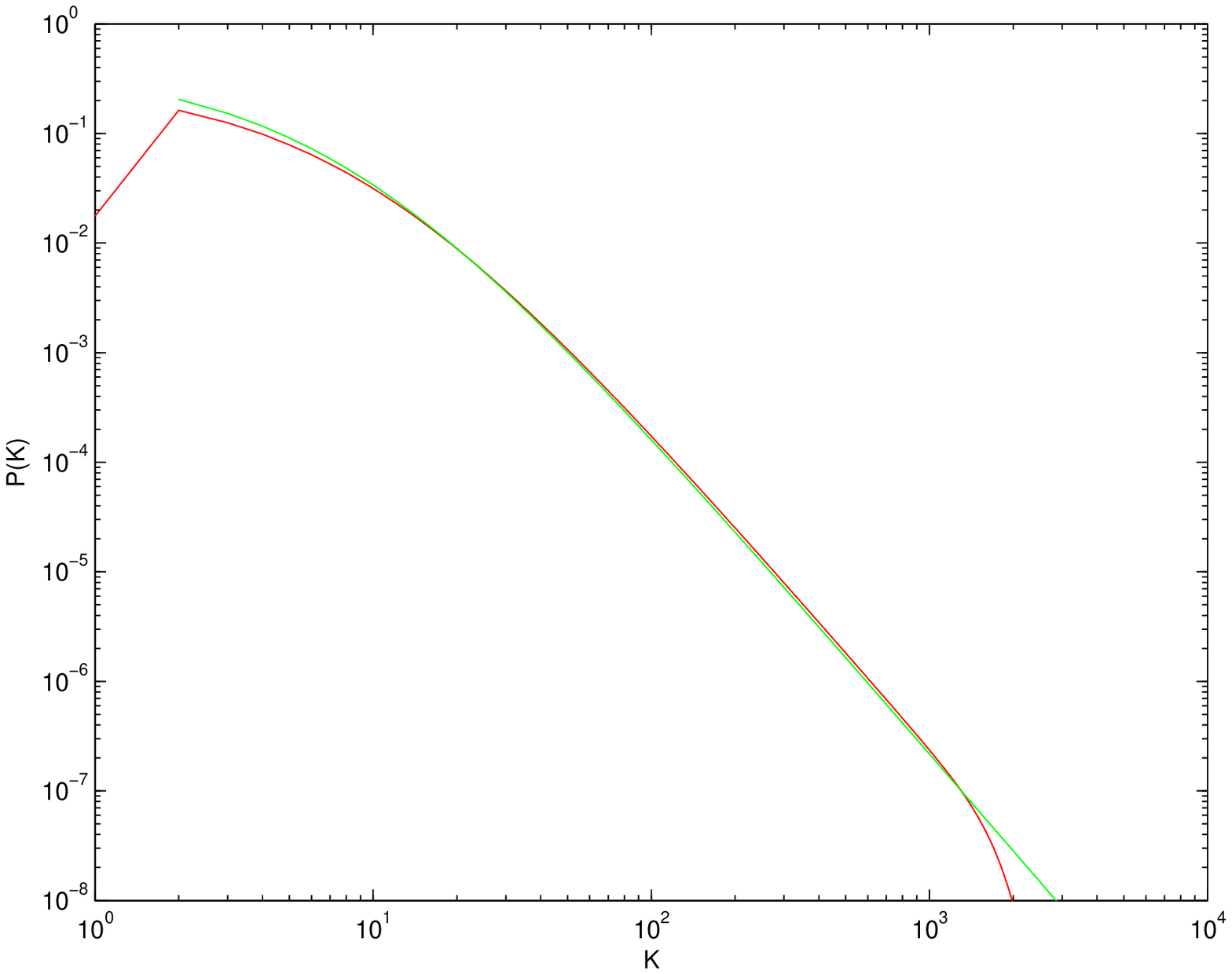}\\
 \caption{{\footnotesize The degree distribution obtained by 
 (\ref{eq: abp})
in solid green line and by the BDP method in solid red line for
$m=2$, $p=0.6$, $q=0.1$, $S=2$ and $t=100000$.}}\label{fig 3}
\end{center}
\end{figure}
\begin{figure}[htp]
 \begin{center}
  \includegraphics[width=6cm]{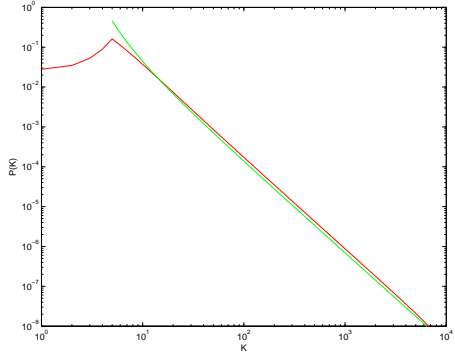}\\
 \caption{{\footnotesize The degree distribution obtained by 
(\ref{eq: abp}) in solid green line and by the BDP method in solid
red line
 for $m=5$, $p=0.2$, $q=0.4$, $S=4$ and $t=150000$.}}\label{fig 4}
\end{center}
\end{figure}

The results from (\ref{eq: abp}) and the BDP method are compared
in figures \ref{fig 3} and \ref{fig 4}. The distribution curves of
the two methods match very well in the middle section.
Again, the distribution curves from the BDP method are {\it horse head} like with a downward bending head. 
(\ref{eq: abp}) over estimates small degree probabilities and does
not provide the probabilities for degrees smaller than $m$.

\newpage

\bigskip

We summarize the results and findings in this paper as follows:
(1) We introduce a simple yet
flexible model of evolving networks with both addition and removal of links and nodes. 
The removal of both links and isolated nodes is new; 
(2) The connection between an evolving
network and a set of non-homogenous birth-and-death processes
provides an efficient algorithm to numerically calculate the
network degree distribution. With this method, 
we reveal the complete process by which a network evolves into a scale-free state;
(3) With the close match between the numerical results and
simulation results, our birth-death method provides
an efficient and reliable substitution to simulation,
in particular since the existing analytical methods cannot handle more complicated network mechanisms
and the computational requirements of simulation are often too high; 
(4) We find that the method based on the continuum theory is not suitable for
small degree distribution and under estimates large degree
probabilities;
(5) The {\it horse head}-like degree distribution
curves have been observed in a number of real networks, such as
the actor collaborations and word co-occurrences networks (see
figure 1 (d) and (e) in Newman \cite{Newman02}). Using the
birth-and-death process method, we demonstrate that the
distribution curves of growing network are {\it snake head} like
while the distribution curves of evolving networks are {\it horse
head} like; and (6) Degree distributions of evolving networks have
two distinct sections and the maximum probability occurs at degree $m$.

\end{document}